# Fully coherent hard X-ray generation by two-stage of Phase-merging Enhanced Harmonic Generation

Guanglei Wang[1], Weiqing Zhang[1], Xueming Yang[1], Chao Feng[2], Haixiao Deng[2*]

[1]*State Key Laboratory of Molecular Reaction Dynamics, Dalian Institute of Chemical Physics, Chinese Academy of Sciences, Dalian 116023, P. R. China*
[2]*Shanghai Institute of Applied Physics, Chinese Academy of Sciences, Shanghai, 201800, P. R. China*

Cascading stages of seeded free electron laser (FEL) is a promising way to produce fully coherent X-ray radiations. We study a new approach to produce coherent hard X-rays by cascading the recently proposed phase-merging enhanced harmonic generation (PEHG). The scheme consists of one dogleg and two PEHG configurations, which may be one of the leading candidates for the extracted undulator branch in future X-ray FEL facilities. FEL physics studies show that such a scheme is feasible within the present technology and can provide high brightness X-ray radiation pulses with narrow bandwidth and fully coherence, and the radiated peak power at 1 Å wavelength converted from an initial 200 nm seed laser is over 2 GW.

## 1 Introduction

The hard X-ray free electron laser (FEL) era has reached by the successful lasing of self-amplified spontaneous emission (SASE) [1]. However, because the FEL radiation starts from the initial shot noise, the output of SASE typically has poor longitudinal coherence and large shot-to-shot fluctuation. At the same time, as the most promising way for delivering fully coherent FEL radiation pulses in the ultraviolet even soft x-ray regions, seeded FEL schemes are arising worldwide. The world's first seeded FEL user facility FERMI started to deliver extreme ultraviolet pulses to users in 2011 [2], and several other FEL facilities based on seeded configurations are under construction or consideration [3-5].

Generally, in a seeded FEL configuration like high-gain harmonic generation (HGHG), an external coherent seed laser pulse is employed to interact with the electron beam in a short undulator (modulator) to generate sufficient beam energy modulation. The electron beam is then sent through a magnetic chicane (dispersion section) where the energy modulation is converted into density modulation and the micro-bunching on the scale of the optical seed laser wavelength is established. Taking the advantage of that the Fourier transform of the density modulation contains abundant components at high harmonic of the seed, coherent short-wavelength signal that dominants over the beam shot noise can be amplified by a relatively long undulator (radiator) resonant at the interested harmonic of the seed. As expected from the theories, the success of seeding schemes such as HGHG [6-9] and echo-enabled harmonic generation (EEHG) [10-13] lead to the possibility of generating short wavelength radiation pulses with high brilliance and excellent longitudinal coherence.

It is expected that, more scientific opportunities ranging from materials and biomaterials sciences, nano-sciences, plasma physics, molecular and chemistry will emerge, as the fully coherent X-ray FEL sources are exploited. However, currently single-stage seeded FELs are not capable for achieving hard X-ray region, because of limited frequency up-conversion efficiency [14]. Therefore, cascading stages of HGHG and EEHG have been proposed to produce fully coherent X-ray radiations [15-16]. More recently, a novel seeded FEL scheme so-called phase-merging enhanced harmonic generation (PEHG) [17-18] is proposed to generate fully coherent short-wavelength radiations. In PEHG, when the transversely dispersed electrons pass through a transverse gradient undulator (TGU) modulator [19], around the zero-crossing of the seed laser, the electrons of same kinetic energy will merge into a same longitudinal phase due to the transverse-longitudinal coupling, which enables a supreme frequency up-conversion efficiency with a very small energy modulation. Theoretical calculations and numerical simulations indicate that a single-stage PEHG is capable of generating high power soft X-ray radiation with narrow bandwidth close to Fourier-transform-limited. Moreover, it is found that PEHG almost has no response to the beam energy curvatures produced in the acceleration process [20].

These above-mentioned properties make PEHG a promising candidate for short-wavelength FELs. Meanwhile, more and more extracted undulator branches are considered in modern X-ray FEL user facilities, e.g., in SACLA, PAL-XFEL and Swiss-FEL [21-23], which offer an opportunity of large transverse dispersion for a PEHG operation. Under such circumstances, cascading stages of PEHG with "fresh bunch" technique [24] is a straightforward idea for extending to the hard X-ray region. In this paper, we first describe the principle of the cascaded PEHG. Then the detailed considerations on the hard X-ray generation by PEHG schemes are shown. The main issues including the design of the dispersion dogleg, undulator system and the seed laser system are presented.

## 2 Layout of cascading PEHG

PEHG scheme combines a dispersion dogleg and a TGU modulator, which induce a transverse-longitudinal phase space coupling. Firstly, the e-beam is transversely dispersed by the dogleg, then the transversely dispersed electrons pass through the TGU modulator with transverse gradient α, and dimensionless undulator parameter K, around the zero-crossing of the seed laser, the electrons with the same energy will merge into a same longitudinal phase. PEHG holds a great promise for generating fully coherent short-wavelength radiation due to the advantage of phase merging effect, the soft X-ray wavelength range can be reached in a single stage PEHG.

Even the PEHG scheme has a supreme up-conversion efficiency; it's still difficult to achieve 1 Å FEL radiation from the commercially available seed laser. Motivated and stimulated by the recent and great success of two-stage HGHG demonstrations and operations at SDUV-FEL [25-27] and FERMI [28], a straightforward and promising idea is the multi-stage PEHG scheme. Now, we investigate the possibility to produce hard X-ray using a cascaded PEHG scheme. The layout is shown in Figure 1, which consists of a PEHG, a magnetic delay chicane and an HGHG configuration, while the HGHG part is equivalent to a second PEHG by jointly using the transverse dispersion section in the initial.

In Figure 1, the dogleg is utilized to obtain a transverse dispersion in the whole e-beam, then a short seed laser pulse is injected and adjusted in the first TGU modulator (M1) to make sure only the e-beam energy of the tail part is modulated, this e-beam is sent through the dispersion section (DS1), which converts the energy modulation into a density bunching. Meanwhile the head part, called "fresh" part hereafter, also experiences a same process as the tail part in the first stage, in the absence of energy modulation. Then the FEL radiation from the first stage radiator (R1) which uses the e-beam tail part, serves as the seed laser of the second stage. The magnetic chicane between the two stages is used to realize the "fresh bunch" delay and make sure the e-beam can exactly interact with the seed laser in the second stage modulator. Noticing that the fresh part has already been transverse dispersed by the dogleg, so the second stage together with the dogleg can also be considered as a whole PEHG. In the second stage PEHG, considering the stringent beam requirements and practical operation difficulties in the short-wavelength, a normal modulator (M2) and a TGU (M3) is used for the energy modulation and the transverse manipulation of the electron beam, respectively. It has been theoretically demonstrated that such a configuration has a larger efficiency of density modulation and is much more flexible for practical operation [18]. In this configuration, the output from the second radiator (R2), which is tuned at the harmonic of the first radiation, can be extended to even hard X-ray region. In addition, the radiator of the first stage can work at the coherent harmonic generation (CHG) [29-30] or saturation regime, depending on the required seed laser power for the second stage.

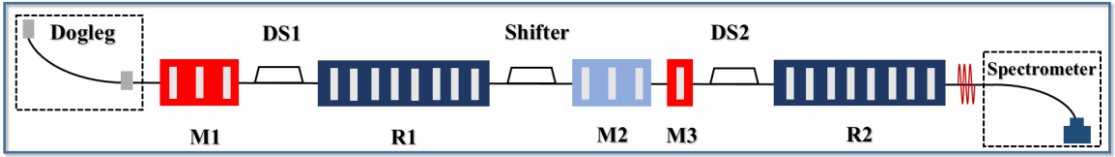

**Figure 1**. The schematic layout of cascaded PEHG to produce hard X-ray.

## 3 Theoretical optimization of cascaded PEHG

If one takes the transverse effects into accounts, the PEHG bunching can be written as:

$$b = J_h(hD\Delta\gamma_s)\exp(\frac{h^2 D \sigma_x^2}{2\eta^2}), \tag{1}$$

where $h$ is the harmonic number, $D = k_s R_{56}/\gamma$, $k_s$ is the wave number of the seed laser, $R_{56}$ is the strength of the dispersive chicane, $\gamma$ is the electrona beam Lorentz factor, $\Delta\gamma_s$ is the energy modulation induced by seed laser and $J_h$ is the $h^{th}$ order Bessel function, $\eta$ is the transverse dispersion of the dogleg and $\sigma_x$ is the transverse beam size. Obviously, the bunching of PEHG is the same as the standard HGHG with an equivalent energy spread of $\sigma_{eff} = \sigma_x/\eta$ [18].

The optimal relationship between the dogleg and the design parameters of TGU is

$$TD = -B, \tag{2}$$

$T = \frac{k_s L_m K_0^2 \alpha \sigma_x}{2\gamma^2}$ is the dimensionless gradient parameter of a TGU with $\alpha$ and $K_0$ represent the transverse gradient and central dimensionless parameter. The reasonable beam size in the TGU modulator is $\sigma_x = \sqrt{\varepsilon_x L_m/2\gamma}$, with $\varepsilon_x$ and $L_m$ for the normalized horizontal emittance and the modulator length, respectively.

According to Eq. (2), the optimized condition of α and η can be summarized in Eq. (3),

$$\eta = -\frac{2\gamma(h+0.81h^{1/3})}{\alpha h k_s L_m K_0^2 \sigma_r} \tag{3}$$

$A$ is the energy modulation amplitude induced by seed laser, $\sigma_r$ is the RMS beam energy spread.

To explore the performance of cascaded PEHG with realistic parameters, we suppose a hard X-ray FEL, using a group of parameters close to the Swiss-FEL, a 6 GeV electron beam with sliced energy spread of 0.6 MeV, i.e., a relative energy spread of $1\times10^{-4}$, normalized emittance of 1.0 μm-rad, and peak current of 3 kA is expected at the exit of the LINAC for efficient FEL lasing. In the FEL section, 1 Å FEL radiation is generated as the $2000^{th}$ harmonic of the initial seed laser. And let's now present the details, we consider a dogleg with $\eta = 1.4$ m for effectively transverse dispersion of the e-beam, then a 200 nm seed laser is imported, the peak power is about 4 GW for sufficient energy modulation, the FEL radiation in each stage will inherent the properties of this high quality fundamental seed laser, and results in a Fourier transform limited FEL radiation pulse. After the two-stage cascaded process, 1 Å FEL radiation is amplified to saturation with a peak power beyond 2 GW by the 40 m long radiator. The parameters for the undulator, the dogleg, and the seed laser system are summarized in Table 1.

Table 1. The main parameters of hard X-ray FEL facility.

|  | The first stage | | The second stage | | |
| --- | --- | --- | --- | --- | --- |
| Undulator | M1 | R1 | M2 | M3 | R2 |
| Undulator period | 0.2m | 0.04m | 0.04m | 0.04m | 0.015m |
| Undulator length | 2m | 25m | 2m | 0.4m | 40m |
| Laser wavelength | 200nm | 4nm | 4nm | | 0.1nm |
| Laser power | 4GW | 1GW | 1GW | | 2GW |
| Laser size | 100mm | 50mm | 100mm | | 50mm |
| Dispersion R56 | 0.105mm | | 1.5μm | | |

For the case of $A = 5$, according to Eq. (1), the PEHG bunching factor at different harmonics is plotted in Figure 2(a), due to the equivalent energy spread compression, the theoretical maximum value at the 50$^{th}$ harmonic is about 18%. Three-dimensional simulation is utilized to illustrate the phase space evolution in the PEHG scheme, the longitudinal phase space after passing through DS1 is shown in Figure 2(b), one can see that the initial beam energy spread is artificially rearranged and suppressed in PEHG scheme by the so-called phase-merging effect, the energy spread compression factor $C = \eta \sigma_r / \gamma \sigma_x$ is about 5 and this compression makes the available harmonic number increase about 5 times with the same bunching factor for high harmonics.

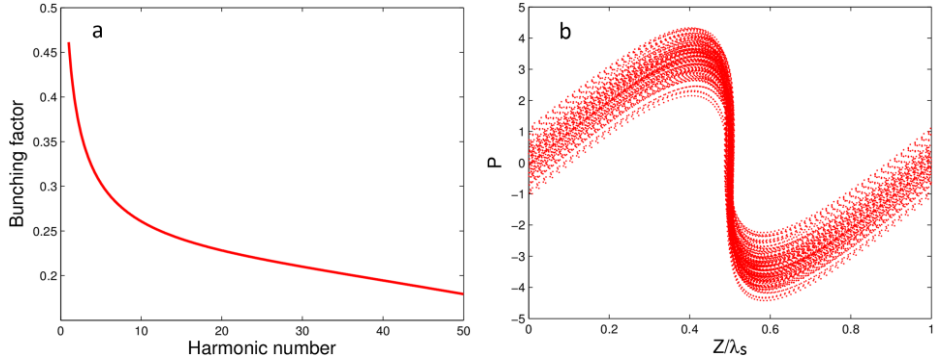

**Figure 2**. Figure 2(a) shows the PEHG bunching factor evolution for the energy modulation amplitude $A = 5$; Figure 2(b) is the PEHG phase space after passing through DS1.

## 4 Hard X-ray based on cascaded PEHG scheme

The parameters of the electron beams, undulators and the seed laser system are summarized in last section. Here we show the details about the FEL operation. The first stage of the cascaded PEHG expects to generate 4 nm FEL radiation from the 200 nm seed laser with longitudinal Gaussian profile, 10 GW peak power. The FWHM length of the seed laser pulse is about 50 fs which is much shorter than the designed electron bunch length. Only a fraction of the beam is modulated in M1 and produce coherent radiation in R1. The radiation will be shifted to a fresh part of the electron by the shifter and severs as the seed laser for the second stage. A drift is located after the first stage for lattice matching and diffusion of the FEL radiation from first stage. A modulator system consists of a normal planar undulator and a TGU is proposed for the second stage PEHG. According to Ref. [18], for a relatively small energy modulation, this configuration enables nearly the same bunching factor compared with the first stage layout, this properties reduce the power requirements for the first radiator and improve the output performances of the cascaded PEHG scheme. Finally, the 1 Å radiation will be generated by the modulated fresh bunch in R2.

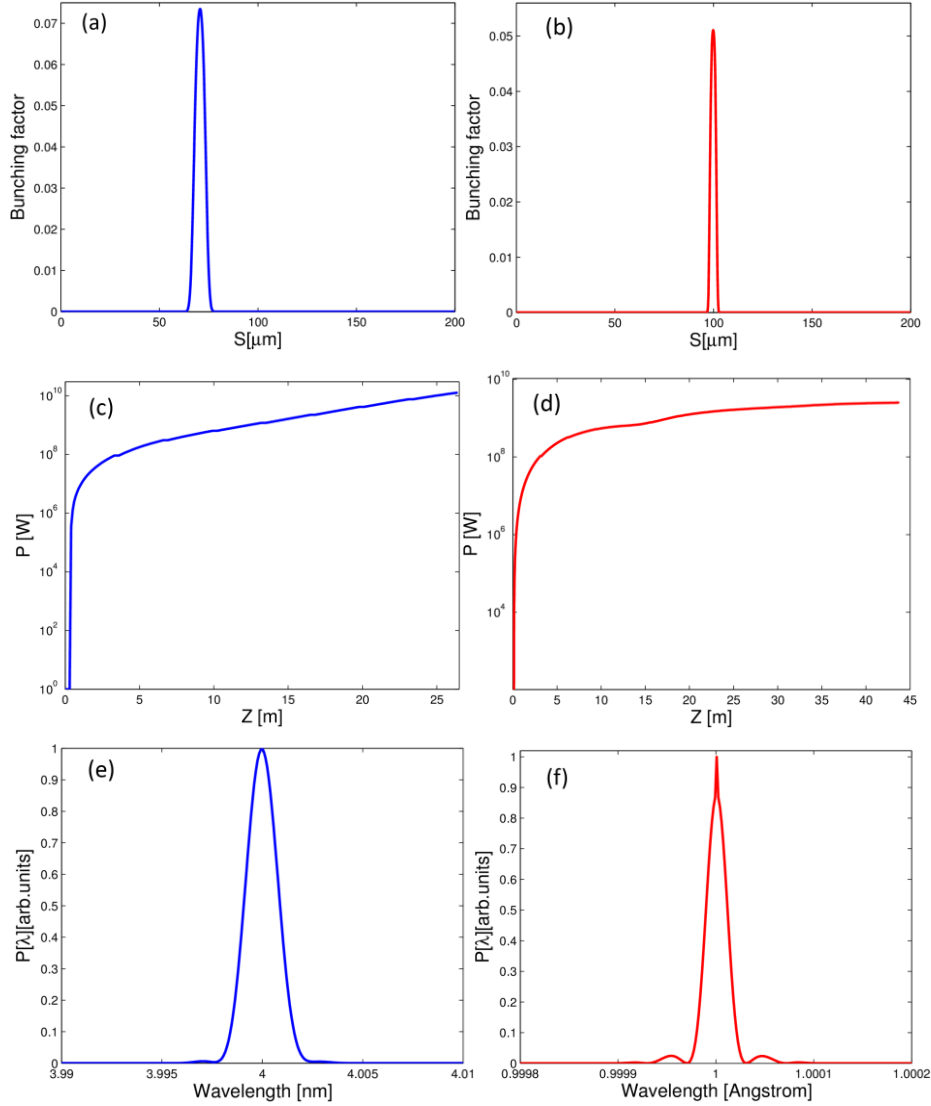

**Figure 3**. FEL performances in two stages. (a) the 50th harmonic bunching factor distribution along the electron beam at the entrance of the radiator in the first stage; (b) the 40th harmonic bunching factor distribution along the electron beam at the entrance of the radiator in the second stage; (c) the output radiation peak power of the first stage; (d) the output radiation peak power of the second stage; (e) the spectrum of the radiation generated by the first stage; and (f) the spectrum of the radiation generated by the second stage.

On the basis of the parameters shown in Table 1, time dependent simulation has been carried out to assess the FEL lasing results by the well-benchmarked FEL code GENESIS [31]. To obtain realistic simulation results, the whole electron beam was tracked through the first stage to the second stage PEHG. The simulation results are illustrated in Figure 3. In view of the tradeoff between the seed laser induced energy spread and the available bunching factor, a moderate energy modulation amplitude of $A_1 = 5$ is chosen for FEL gain process in the M1, then the e-beam is well bunched after the DS1, the bunching factor of 50[th] harmonic is optimized to 7% at the entrance of the first radiator. The 4 nm FEL radiation pulse is generated with the output power of 5 GW and the pulse length of the seed laser is also maintained. A matching section is located after the first stage in order to provide adjustable beta-matching, diffusion of FEL spot and smear out the e-beam micro-bunching generated in the first stage. The energy modulation amplitude for the second stage is $A_2 = 5$ and the 40th harmonic bunching factor is optimized to be 5% (Figure 3 (b)), the FEL radiation generated by the fresh part saturates after 40 m with a peak power of 2 GW. After passing through 40 m long radiator, the relative FWHM bandwidth of the 1 Å radiation is about $4 \times 10^{-5}$ and close to the Fourier transform limited, the spectral emission presents regular quasi perfect Gaussian shape pulse to pulse, little redshift compared with 1 Å radiation is an effect due to the energy loss of the e-beam in the lasing process. The noisy spike and little FEL spectrum broaden are induced mainly by the amplification of intrinsic shot noise in the e-beam, which is proportional to the harmonic number $h$. Moreover, the FEL peak power in Figure 3 (b) and (d) increases continuously, because the main FEL pulse slips to the unsaturated electron beam in the radiator.

## 5  Dogleg design for the cascaded PEHG beam line

The PEHG mechanism requires a large dogleg to provide the $R_{16}$ used for transverse dispersion of the e-beam, then this necessary dispersion will lead to increased beam size and transverse emittance, definitely degrade the PEHG performance. Thus, we concentrate on the effects of the dispersion dogleg in e-beam in this section. Two dipole doglegs are widely used to translate the beam axis horizontally or vertically, quadrupoles are placed between the two consecutive dipoles to match first-order dispersion and provide beta focusing. The transform matrix of a dogleg can be described as Eq. (4), where $L$ is the length of drift space, $\theta$ and $\rho$ is the bend angle and bend radius, respectively.

$$\begin{vmatrix} 1+\frac{L}{\rho}\cos\theta\sin\theta & L\cos^2\theta & 0 & 0 & 0 & -L\cos\theta\sin\theta \\ \frac{L}{\rho^2}\sin^2\theta & 1-\frac{L}{\rho}\cos\theta\sin\theta & 0 & 0 & 0 & \frac{L}{\rho}\sin^2\theta \\ 0 & 0 & 1 & L & 0 & 0 \\ 0 & 0 & 0 & 1 & 0 & 0 \\ -\frac{L}{\rho}\sin^2\theta & -L\cos\theta\sin\theta & 0 & 0 & 1 & L\sin^2\theta \\ 0 & 0 & 0 & 0 & 0 & 1 \end{vmatrix} \quad (4)$$

The dispersion strength of a dogleg can be estimated as $\eta = -Lcos\theta sin\theta$, and as mentioned, the $\eta$ is about $1.4\ m$ for the cascaded PEHG scheme to generate sufficient transverse off-set in e-beam. In this section, we design two different doglegs to compare the effects of transverse dispersion in e-beam, the corresponding bend angle of this dogleg is designed as 3 degree. A so-called $\pi$ insertion composed of quadrupoles will be placed between the bending magnets to cancel the effects of elements placed on either side of them. For an achromatic dogleg, a same configuration is utilized for comparison, the beta function at the exit of the two schemes is nearly the same, both are about 25 m. ELEGANT is used for the simulation of beam evolution in the dogleg [32]. Figure 4 shows the e-beam transverse phase space distribution after the dogleg, due to the dispersion, an obvious energy chirp happened along the $\sigma_x$ coordinate of the e-beam, this artificially energy chirp is adjustable to match the transverse gradient of the TGU magnet field. The beam size is about 3 times increased for the dispersion dogleg, what's more, the dispersion terms will introduce a transverse position jitter related with the e-beam energy jitter, so a large seed laser transverse spot size is required to cover the whole e-beam, which means that high pulse energy seed laser is needed.

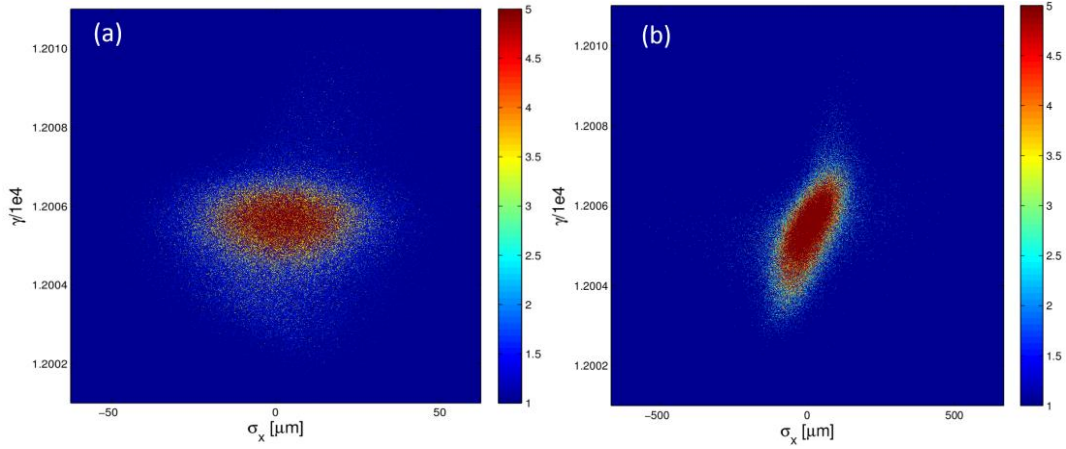

Figure 4: the e-beam phase space distribution after achromatic dogleg (a) and dispersive dogleg (b), the bend angle of the dogleg is 3 degree and the total length is about 29 m.

Figure 5(a)-(b) show the lattice and dispersion evolution along the dogleg coordinate, the beta function at the exit is nearly the same, both are about 25 m, to optimize the beam size after the transport process; the corresponding dispersion is summarized in figure 5(b), the parameter $\eta$ can be controlled by the π insertion effectively. Figure 5 (c) and (d) show the simulation results of the e-beam at the exit of the dogleg, due to the increase of transverse beam size in x-direction, the transverse emittance is almost two times increase compared with the achromatic case. Benefit from the small deflection angle, the effects of coherent synchrotron radiation (CSR) in emittance increase can be ignored in our case. The emittance increase will definitely degrade the PEHG bunching performance, which however can be compensated with large energy modulation amplitude [17-18]. In Figure 5 (c), due to the wake field and un-vanished $R_{56}$ terms, a little increase of bunch length can be observed, the double horn distribution is mainly induced by the x-z coupling happened in e-beam and the statistical methods of ELEGANT itself. As mentioned above, the dogleg is in horizontal, and the beam quality in y-direction can be perfectly preserved. Finally, it's worth emphasizing that the beam tracking in this section is based on an ideal e-beam, more detailed issues, such as the effects of the strong CSR in dipole, beam lattice evolution in the undulator system and effects of beam distribution along the time coordinate should be further studied.

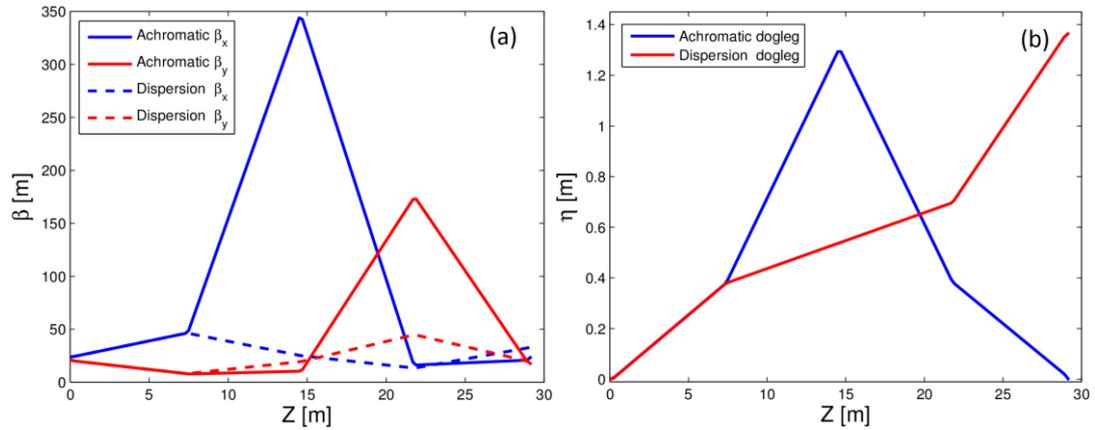

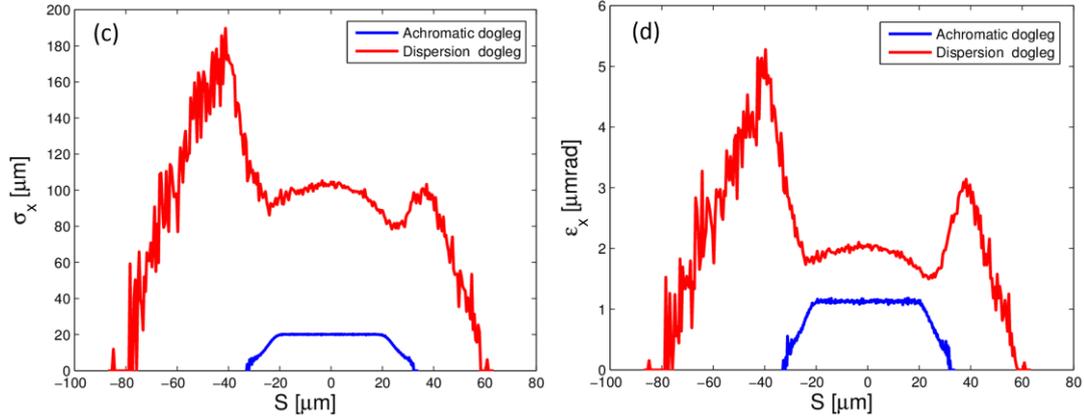

Figure 5: Beta function (a) and dispersion evolution (b) along the dogleg coordinate; transverse e-beam size in x-direction (c) and emittance growth due to the dispersion of dogleg (d), with (red) and without dispersion (blue).

# 6 Conclusion

The design studies of cascaded PEHG scheme for potential hard X-ray FEL projects are presented. Compared with the cascaded HGHG, only two-stage PEHG are needed for achieving hard X-ray region in this scheme. Due to the living doglegs in multi-branches FEL facility, the cascaded PEHG is simple in layout, and has easier laser-beam synchronization and better FEL spectrum in real operation than the cascaded EEHG. The results here show that using the realistic beam parameters, 0.1 nm coherent hard X-ray FEL with peak power up to 2 GW can be generated directly from a 200 nm seed laser based on this scheme. Some practical limiting factors that might affect the performance of cascaded PEHG scheme are also taken into account. According to the simulation, it is also found that the dogleg will introduce an increase of the horizontal beam size, and it is necessary to utilize a powerful seed laser to modulate the e-beam, these requirements can be totally fulfilled by the state-of-the-art laser technology. Further more detailed studies, such as the CSR in the bending magnet, e-beam lattice in the undulator system and the effects of the longitudinal beam distribution are still needed to be analyzed in the future.

# Acknowledgement

The authors are grateful to Bo Liu, Dong Wang and Zhentang Zhao for helpful discussions. This work was supported by the National Natural Science Foundation of China (21127902 & 11322550) and Ten Thousand Talent Program.

# References

[1] Emma, P. et al, Nature Photonics, **4:** 641-647 (2010)

[2] E. Allaria et al, Nature Photonics, **6:** 699-704 (2012)

[3] http://flash2.desy.de/

[4] Concept Design Report, Soft X-ray Free Electron Laser Test Facility, Shanghai, May 2015

[5] Deng, H. et al, Chin. Phys. C, **38:** 028101 (2014)

[6] L. H. Yu, Phys. Rev. A, **44:** 5178 (1991)

[7] L. H. Yu, et al, Science, **289:** 932-934 (2000)

[8] L. H. Yu, et al, Phys. Rev. Lett., **91:** 074801 (2003)
[9] M. Labat, et al, Phys. Rev. Lett., **107:** 224801 (2011)
[10] G. Stupakov, Phys. Rev. Lett., **102:** 074801 (2009)
[11] D. Xiang, G. Stupakov, Phys. Rev. ST Accel. Beams, **12:** 030702 (2009)
[12] D. Xiang, et al, Phys. Rev. Lett., **105:** 114801 (2010).
[13] Z. T. Zhao, D. Wang, et al, Nature Photonics, **6:** 360-363 (2012)
[14] E.L. Saldin, et al, Opt. Commun., **202**: 169-187 (2002)
[15] Wu J H, et al, Nucl. Instrum. Methods Phys. Res., Sect. A, **475:** 104-111 (2001)
[16] C. Feng, et al, Chin. Sci. Bull., **55:** 221-227 (2010)
[17] H. X. Deng, C. Feng, Phys. Rev. Lett., **111:** 084801 (2013)
[18] C. Feng, H. Deng, D. Wang, Z. Zhao, New J. Phys., **16:** 043021 (2014)
[19] T. I. Smith et al, Journal of Applied Physics, **50:** 4580 (1979)
[20] G. Wang, et al, Nucl. Instrum. Methods Phys. Res., Sect. A, **753:** 56-60 (2014)
[21] J. H. Han, et al, TUPPP061, Proceedings of IPAC2012, New Orleans, Louisiana, USA
[22] Makina Yabashi, et al, J. Synchrotron Rad., **22:** 477-484 (2015)
[23] SWISS-FEL Concept Design Report.
[24] Yu L H, Ben-Zvi I, Nucl. Instrum. Methods Phys. Res., Sect. A, **393:** 96-99 (1997)
[25] Deng H X, Dai Z M, Chin. Phys. C, **32:** 236-242 (2008)
[26] Feng C, et al, Chin. Sci. Bull., **57:** 3423-3429 (2012)
[27] B. Liu, et al, Phys. Rev. ST Accel. Beams, **16:** 020704 (2013)
[28] E. Allaria et al, Nature Photonics, **7:** 913-918 (2013)
[29] Coisson R, MartiniF D, Phys Quant Electron, **9:** 939-960 (1982)
[30] Goloviznin V, Amersfoort P W, Phys. Rev. E, **55:** 6002-6010 (1997)
[31] S. Reiche, Nucl. Instr. and Meth. A, **429:** 243-248 (1999)
[32] M. Borland, ANL Advanced Photon Source, Report No. LS-287, 2000